\documentclass[12pt,preprint]{aastex}

\shorttitle{The CaT of Elliptical Galaxies}
\shortauthors{Saglia et al.}

\begin{document}


\title{The puzzlingly small CaII triplet absorption in elliptical galaxies}


\author{R.P.~Saglia, 
Claudia Maraston, Daniel Thomas, Ralf Bender,}
\affil{Universit\"ats-Sternwarte,
Scheinerstra\ss e 1, D-81679 Munich, Germany and\\ Max-Planck-Institut
f\"ur extraterrestrische Physik, Postfach 1312, D-85741 Garching,
Germany; saglia@usm.uni-muenchen.de, maraston@usm.uni-muenchen.de,
daniel@usm.uni-muenchen.de, bender@usm.uni-muenchen.de}
\and
\author{Matthew Colless}
\affil{Research School of Astronomy and Astrophysics, The Australian National
University, Weston Creek, ACT 2611, Australia; colless@mso.anu.edu.au}




\begin{abstract}
We measure the central values (within $R_e/8$) of the CaII triplet
line indices CaT$^*$ and CaT and the Paschen index PaT at 8600 \AA\
for a 93\%-complete sample of 75 nearby early-type galaxies with
$B_T<12$ and $V_{gal}<2490$.  We find that the values of CaT$^*$ are
constant to within 5\% over the range of central velocity dispersions
$100\le \sigma\le 340$ km/s, while the PaT (and CaT) values are mildly
anti-correlated with $\sigma$.  Using simple and composite stellar
population models, we show that: a) The measured CaT$^*$ and CaT are
lower than expected from simple stellar population models (SSPs) with
Salpeter initial mass functions (IMFs) and with metallicities and ages
derived from optical Lick (Fe, Mg and H$\beta$) indices. Uncertainties
in the calibration, the fitting functions and the SSP
modeling taken separately cannot explain the discrepancy. On the average, the
observed PaT values are within the range allowed by the models and the
large uncertainties in the fitting functions. 
b) The steepening of the IMF at low masses
required to lower the CaT$^*$ and CaT indices to the observed values
is incompatible with the measured FeH index at 9916 \AA\ and the
dynamical mass-to-light ratios of ellipticals. c) Composite stellar
populations with a low-metallicity component reduce the
disagreement, but rather artificial metallicity distributions are
needed. Another explanation may be that calcium is indeed underabundant in
ellipticals. 
\end{abstract}




\section{Introduction}

\label{introduction}

The determination of the mean ages and metallicities of local
elliptical galaxies is one of the key observational tests of models
for galaxy formation and evolution. The classical picture of
monolithic collapse (Larson 1974) assumes high formation redshifts and
passive evolution, producing large central metallicities with strong
gradients. In contrast, semi-analytic models of galaxy formation
embedded in the hierarchical structure formation typical of Cold Dark
Matter universes (Kauffmann, 1996), produce elliptical galaxies
through mergers of disks and predict a large spread in the formation
ages (especially for field ellipticals), with solar mean metallicities
and shallow gradients.  Although many indications support the merging
scenario, a clear-cut answer from the observations has been hampered
by the age-metallicity degeneracy in the spectra of (simple) stellar
populations: the same broad-band colors and absorption features can be
obtained for very different combinations of ages and
metallicities. The combined use of indices more sensitive to age (like
the Balmer lines) and those more sensitive to
metallicity (like Mg$_2$ and Fe5270 and Fe5335) offers a way out
(Worthey 1994). However ambiguities remain, because the presence of a
small metal-poor old stellar population may bias the age estimate
(Maraston \& Thomas, 2000, MT), and the metallicity estimate based on
the Mg indices are systematically higher than the ones using iron
lines (the so-called Mg over Fe over-abundance problem; Worthey, Faber
\& Gonz\'alez, 1992, Trager et al. 2000, Thomas, Maraston \& Bender
2002, TMB).  Recent modeling of the CaII triplet line 
at 8600\AA\ (Idiart, Th\'evenin \& De Freitas Pacheco 1997, I97;
Garc\'{\i}a-Vargas, Molla \& Bressan 1998; Moll{\'a} \&
Garc\'{\i}a-Vargas, 2000) concluded that this index is insensitive
to age for populations older than 1 Gyr, raising hopes that it
could allow a robust measure of metallicity, and, in combination with
colors or indices, age. An excellent review of the literature on the
subject can be found in Cenarro et al. (2001a, C01).

Here we present the results of our survey of the CaII triplet line in local
elliptical galaxies in the light of the new definition of the index
given by C01 and their accurate determination of
its stellar calibrators (the so called fitting functions, FF, 
Cenarro et al. 2001b, Cenarro et al. 2002, C02). In \S
\ref{observations} we describe the observations and the data
reduction, in \S \ref{results} we present the data, discuss them
with new stellar population models and draw our conclusions.

\section{Observations and data reduction}
\label{observations}

We observed the CaII triplet region at 8600 \AA\ along the major axis
of 94 early-type ellipticals from the Faber et al. (1989) catalog
(hereafter F89). The results presented in this paper are based on the
subsample with $B<12$ mag and $V_{gal}<2490$ km/s, where we observed
75 galaxies, reaching 93\% completeness.  At $V_{gal}\approx 2500$
km/s the reddest window of the indices considered here moves into a
region of strong sky emission lines, making the measurements sensitive
to sky subtraction errors.

The observations were performed at the 3.5m Calar Alto in February and
April 2001, the La Silla ESO NTT in May and November 2001, and the
Siding Spring 2.3m telescope in May and October 2001 and January 2002,
with instrumental resolution ranging from 70 to 80 km/s. Details will
be presented in Saglia et al. (in preparation).  Typically, two
spectra along the major axis of the galaxy were obtained with summed
exposure times ranging between 1 and 3~h.  Template stars used in the
determinations of the fitting functions of C01 were also observed,
trailed along the slit.  The standard CCD data reduction was carried
out under the image processing package MIDAS provided by ESO. A
detailed description of the procedure can be found in Saglia et
al. (in preparation). The wavelength calibration procedure achieved
0.1 \AA\ rms precision. The systematic residuals after sky
subtraction, estimated from the blank sky frames, are less than 1\%.

The galaxy kinematics were derived using the Fourier Correlation
Quotient (FCQ) method (Bender 1990) following Bender, Saglia and
Gerhard (1994). The line indices were measured following the
prescriptions of C01 for the computation of the
newly defined ``generic'' indices CaT, PaT, and CaT$^*=$CaT-0.93PaT. 
In summary, generic indices have an arbitrary number of continuum and
spectral features bandpasses and the pseudo-continuum is derived by
using an error-weighted least-squares fit to all the pixels of the
continuum bands. The CaT index measures the combined strength of the
Ca1, Ca2 and Ca3 lines similar to Diaz, Terlevich \& Terlevich (1989)
and I97, and has a possible contribution for the Paschen lines P16,
P15 and P13. The PaT index measures the combined strength of the
Paschen lines P17, P14 and P12. The CaT$^*$ measures the strength of
the CaII triplet corrected for the contamination from Paschen lines.
At our resolution (see below) we found no significant difference in
the measurements, if the flux calibration (performed by C01) was
applied or not. 
 
The statistical errors were derived from Monte Carlo simulations,
which also take into account the uncertainties from the kinematics
through the redshift and the broadening correction. The errors from
systematics in the sky subtraction (in general at the 1\% level, see
above) were found to dominate in the outer parts of the galaxies.  The
procedure described in Mehlert et al. (1998) was followed to achieve a
uniform focus with wavelength, when necessary.  A correction for the
broadening of the lines due to the internal kinematics of the galaxies
was applied to scale the indices to the instrumental resolution, using
a K-type stellar template. As discussed by C01, this correction is
spectral type dependent and therefore somewhat uncertain; it is small
($\le 5$\%) for the CaT and CaT$^*$ indices, but as large as 20\% at
$\sigma\approx 300$ km/s for the PaT index. The uncertainties in the
correction are $\approx 0.1$ \AA . Finally, since the C01 system is
based on stellar spectra taken with 22 km/s resolution, nearly a
factor 4 higher than the present observations, the dataset was
calibrated on this system by comparing the values of the indices for
the template stars in common and applying linear corrections to the
CaT and PaT indices. Central values of the velocity dispersion
$\sigma$ and of the indices were derived by averaging the profiles
within $R_e/8$ (with $R_e$ taken from F89), luminosity-weighting the
datapoints. The typical statistical errors on the central indices are
smaller than 0.1 \AA . The rms of the differences of the central
values of the galaxy
repeats is 0.2 \AA , having applied a 0.3 \AA\ correction to one of
the runs.

\section{Results and Discussion}
\label{results}

\begin{figure}
\includegraphics[scale=.80,angle=0,origin=c]{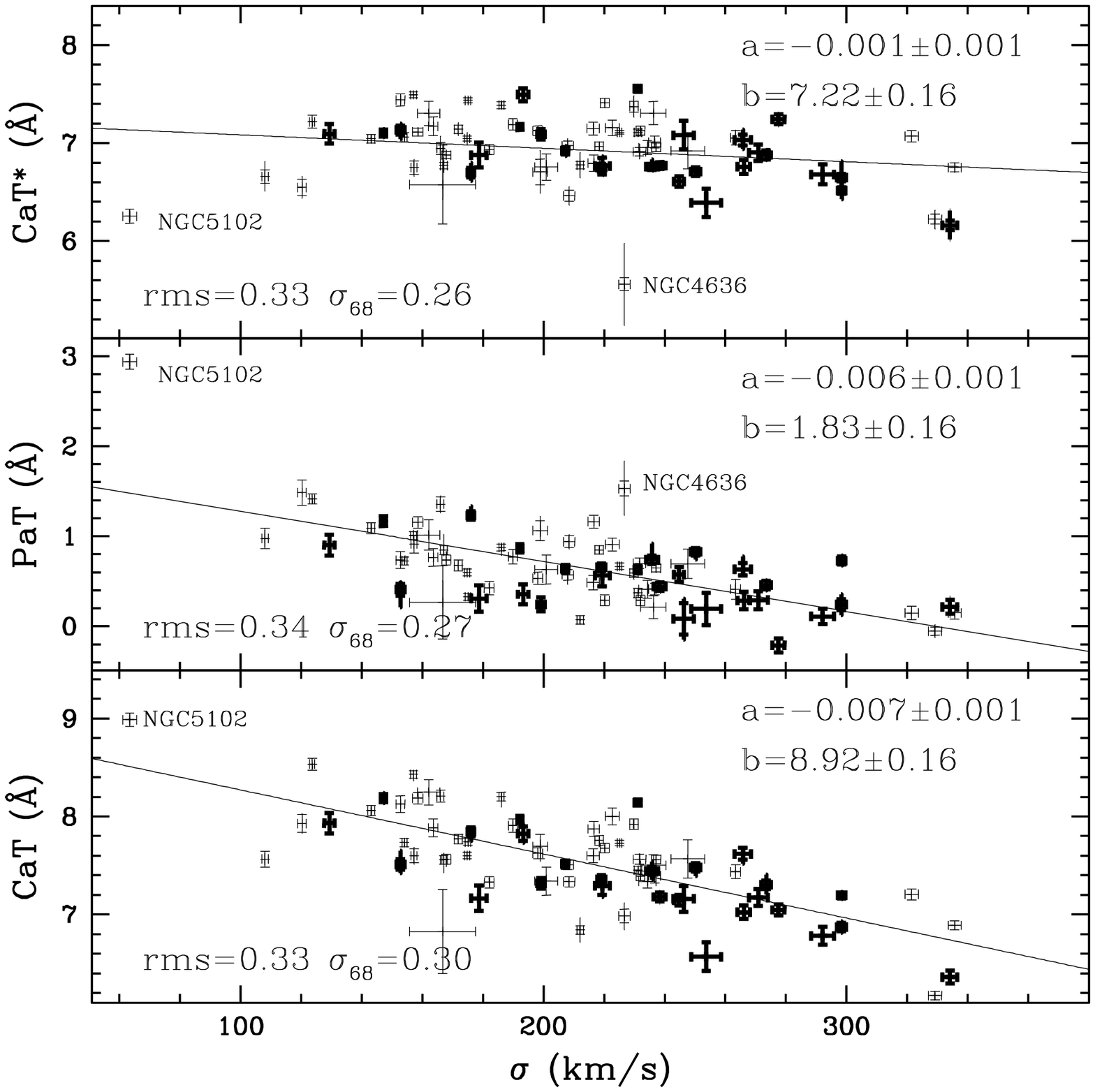}
\caption{The relation between the central velocity dispersion
(averaged within $R_e/8$) and the CaT$^*$, PaT and CaT indices. The
galaxies appearing also in Fig. \ref{figssp}) are shown thick.  Error
bars show the statistical errors. Systematic errors due to sky
subtraction of $\pm 1$\% exceeding the statistical errors are
indicated prolonging the error bars beyond the horizontal mark. The
two most deviant galaxies are labelled. The lines show least-square
fits to the data points. Labels give their slopes ($a$) and
zero-points ($b$), the derived errors, the rms about the best fits and
its robust determination (at 68\% of the cumulative distribution of
the residuals).
\label{figcatsig}
}
\end{figure}

Fig. \ref{figcatsig} shows the relation between the CaT$^*$, PaT and
CaT indices as a function of the central velocity dispersion. As
already noted by Cohen (1979), Faber and French (1980) and Terlevich,
Diaz \& Terlevich (1990), elliptical galaxies have very similar
central values of Calcium triplet index. Averaged over the galaxy
sample, the CaT$^*$ has a mean of 6.93\AA\ and rms 0.33 \AA , or
$\approx 5$\%, just above the measurement errors (statistical,
systematic and due to calibration). Within the derived errors, the
CaT$^*$ index does not depend on $\sigma$, while a mild
anticorrelation is observed for both PaT and CaT, driven by the slightly
larger PaT at lower sigmas. This contrasts
with the behaviour of the Mg$_2$ and Mg$b$ line indices, known to
correlate strongly with $\sigma$ in elliptical galaxies (Bender,
Burstein \& Faber 1993, Colless et al. 1999). These indices trace the
$\alpha$-element magnesium (Tripicco \& Bell 1995, Maraston et
al. 2002), and if the CaII triplet indices were to trace the calcium
abundance, also an $\alpha$-element, a correlation with
$\sigma$ could have been expected.

\begin{figure}
\includegraphics[scale=.59,angle=-90,origin=c]{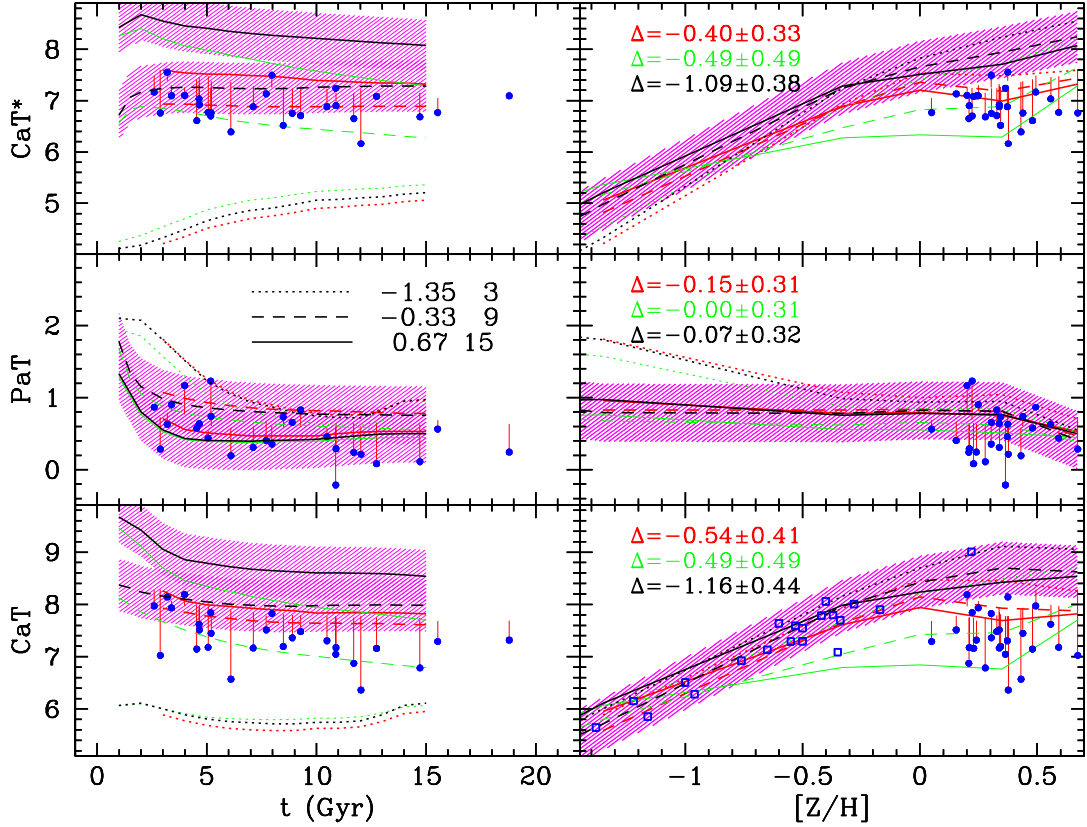}
\caption{Stellar population models of the CaT$^*$, PaT
and CaT indices as a function of age (left panel) and metallicity
(right panel). The black lines show simple stellar populations (SSPs)
with Salpeter ($\alpha=-2.35$) IMF. The magenta shaded areas centered
on the [Z/H]=0.67, -1.35 and t=9 lines show the range  
allowed by the uncertainties in the FFs. 
The green lines show SSPs with IMF slope
$\alpha=-4$ for $M<0.6M_\odot$ and $\alpha=-2.35$ at larger masses. The
red lines show composite stellar population models, where 10\% of the
mass has low metallicity ([Z/H]$=-2.25$). The different line types show
different metallicities (left panel) and ages (right panel), as given
in the key in the left middle panel. The blue
dots show the galaxy sample, with ages and metallicities determined
from Lick indices (see text). The vertical red bars point to the
predicted values of the composite stellar populations (CSPs) models
with a low-metallicity component. 
The mean differences $\Delta$ and
rms between measured and predicted values are given color-coded as
above in the labels.}
\label{figssp} 
\end{figure}

Fig. \ref{figssp} shows stellar population models of the CaT$^*$, PaT
and CaT indices constructed using the FF subroutines of C02 and the
updated code of Maraston (1998, M98). A detailed description of the
models considered here will be given in Maraston et al. (in
preparation). The black lines show simple stellar population (SSP)
models with the Salpeter IMF as a function of age and
metallicity. These models reproduce well the tight metallicity-CaT
correlation observed for globular clusters (open blue squares, from
Armandroff and Zinn 1988, transformed using the relation given by C01,
[Z/H] from Harris 1996). At metallicities higher than solar the models
flatten and the age dependence becomes more important. The PaT index
varies strongly with age for ages less than a few Gyr.  It reaches a
value of $\approx 1$ \AA\ at high ages. Note, however, the large rms
uncertainties of the FFs ($\approx 0.4$ \AA).

The blue filled circles show the subsample of the database
investigated here where ages and metallicity estimates are available
from the analysis of the Fe, Mg and H$\beta$ Lick indices (TMB). 
The results discussed in the following do
not change if the set of ages and metallicities of Terlevich \& Forbes
(2002) are used instead. The same applies to the effects of errors on
the age, metallicity and index, explored using Monte Carlo
simulations. Within the large uncertainties allowed by the FFs, the
models reproduce the average value of the measured PaT indices.  

In contrast, the models predict values of CaT$^*$ and CaT more than 1
\AA\ larger than the measured ones. Such a large discrepancy cannot be
explained by calibration errors, which are at least a factor 5
smaller. Uncertainties in the FFs (rms$\approx 0.5$\AA) alone seem
also unable to explain it, although at high
metallicities ([Z/H]$>0.3$) the FFs are based on only a handful of
stars. The CaT$^*$ of a high Z SSP is dominated by the contribution of the
RGB and the red clump (RC). If we set artificially the value of the CaT$^*$
FF at [Z/H]$=0.35$ from the giant branch phase on to 8 \AA\ (the
lowest value measured for stars in this phase, most of the stars have
CaT$^*=9-10$ \AA ), we obtain CaT$^*=7.5$ \AA\ for the SSP, still 0.5
\AA\ larger than what observed in ellipticals. The flux contribution
of the RC can vary within $\approx 30$\%, due to the
uncertainties on the lifetime of this phase (Zoccali et al. 2000a).  If
we reduce the flux of the RC by this amount, we again obtain
CaT$^*=7.5$ \AA\ for the SSP. Only a 50\% reduction would produce
CaT$^*\approx7$ \AA .

The SSP Salpeter models translate the averaged observed value of the
CaT$^*$ into [Z/H]$=-0.5\pm 0.1$, or Z=$0.3\pm 0.1$ Z$_\odot$. Taking
this result at face value, if the CaT$^*$ traces the calcium
abundance, this could indicate that this element is underabundant, a
suggestion already put forward by Peletier et al. (1999) considering 3
ellipticals. A similar effect has been suggested by McWilliam and Rich
(1994) and Rich and McWilliam (2000), who analyse a sample of metal
rich bulge stars, suggesting that while Mg and Ti are overabundant
with respect to Fe, Ca and Si have nearly solar ratios. In addition,
modeling the Ca4227 Lick index TMB find [Ca/Mg]$\approx -0.3$ in
ellipticals (see also Vazdekis et al. 1997).  However, the issue of Ca
underabundance in high-metallicity stars is still controversial
(McWilliam 1997); C02 point out that the CaT index does not correlate
with the [Ca/Fe] stellar overabundance; and current modeling of the
yields of Type II supernovae (Woosley \& Weaver 1995) does not allow
much room in this direction. While Mg (and O) are produced in the
greatest quantities in high mass stars ($\approx 35~M_\odot$), lower
mass stars ($\approx 15-25~M_\odot$) are responsible for the
production of Ca (and Si). An IMF biased to the high masses or
extremely short ($\approx 10^7$ yr) star-formation bursts that avoid
the Ca enrichment at high metallicities, as discussed by M\'olla and
Garc\i a-Vargas (2000), do not seem a very appealing solution.

An alternative explanation of the low value of the CaT$^*$ index could
be a steeper IMF at low stellar masses, since the CaT$^*$ index decreases
with increasing gravity for cool stars.
The green lines of Fig. \ref{figssp} show SSP models with
IMF slope $\alpha=-4$ for $M<0.6M_\odot$ and $\alpha=-2.35$ (Salpeter) at
larger masses. The CaT$^*$ and CaT indices at the high metallicities
have a stronger dependence on age and are indeed able to reproduce the
measured values of ellipticals. However, the observational evidence
points to an IMF flatter than Salpeter in the bulge of our Galaxy
(Zoccali et al. 2000b). In addition, as already discussed by
Carter, Visvanathan and Pickles (1986) and Couture and Hardy (1993),
such dwarf-dominated models produce values of the FeH feature at 9916
\AA\ that are an order of magnitude larger than what
is observed. Moreover, the visual mass-to-light ratios (at t=10 Gyr,
Z=Z$_\odot$, one gets $M/L_B=40~M_\odot/L_\odot$, M98) are
incompatible with dynamical estimates (Gerhard et al. 2001,
$M/L_B\approx 6$).

Since at low metallicity the CaT$^*$ and CaT indices depend strongly
on Z (see Fig. \ref{figssp}), Composite Stellar Populations (CSPs)
with a low metallicity component are bound to produce (significantly)
lower CaT$^*$ and CaT indices than SSPs with the same (high) mean
metallicity, while leaving the PaT values essentially unchanged.  The
red lines of Fig. \ref{figssp} explore this case, presenting CSPs
where 90\% of the mass is an SSP model and 10\% is the [Z/H]=-2.25 SSP
with the same age, as done in MT. They show that this admittedly
rather artificial model is compatible with the UV constraints
avaliable for ellipticals, generating at the same time higher values
of the H$\beta$ index than SSPs of the same metallicity. The CSP
models of Fig. \ref{figssp} predict almost constant values of the
CaT$^*$ and CaT indices for Z$\ge$ Z$_\odot$. This stems from the
increasing relative importance of the flux at 8600 \AA\ of
the low metallicity component with increasing metallicity of the main
SSP component. It becomes largest at Z$\approx$ Z$_\odot$, to decrease
slightly at higher metallicities, where more flux is emitted at the
near-infrared wavelengths (see M98). Models with broader metallicity
distributions (for example, the closed-box one) smear out this effect,
resulting in CaT$^*$ and CaT indices steadily increasing with
metallicity. The CSP models match the upper third of the galaxy
distribution, however on average they produce CaT$^*$ and CaT indices
still $\approx 0.5$ \AA\ larger than the observed ones. As discussed
in MT, a low metallicity tail is expected in the projected line of
sight metallicity distribution, if radial metallicity gradients are
present. This is suggested by the analysis of the Lick indices
(Davies, Sadler \& Peletier 1993, Mehlert et al. 2000) and color
gradients (Saglia et al. 2000). A ``halo-like'' low-metallicity
population could also be in place, reminiscent of the bimodal (color)
distributions observed in the globular cluster systems of many giant
ellipticals (Larsen et al. 2001 and references therein). 
Finally, we note that CSP models
with a young, metal-rich component, postulated to explain the observed
high central H$\beta$ values of some ellipticals (De Jong \& Davies
1997) do not help here, since at high metallicities the SSPs predict
increasing values of the CaT$^*$ and CaT indices with decreasing ages.
  
To conclude, none of the discussed options alone seems able to explain
the observed distribution of the calcium triplet values and a
combination of them might be at work. In particular, it could be that Ca is
in fact underabundant in elliptical galaxies.

\acknowledgments

RPS acknowledges the International Research EXchange fellowship that
made possible his visit to the Research School of Astronomy and
Astrophysics of the
Australian National University in Canberra, where part of the research
project described here was performed.  RPS, RB, CM and DT acknowledge
the support by the DFG grant SFB 375.  We thank S. Edwards,
C. Harrison and L. Pittroff for helping during the observations and
data reduction. We acknowledge the discussion with Laura Greggio.  We
acknowledge the observatories that supported this project: Calar Alto
(Centro Astrofisico Hispano Alemano), Siding Spring (MSSSO) and La
Silla (ESO).




\clearpage


\clearpage 

\end{document}